\documentclass[a4,11pt]{reportN}
 
\usepackage{icomma} 

\usepackage[brazil]{babel}    
\usepackage[utf8]{inputenc}
\usepackage{graphics}
\usepackage{subfigure}
\usepackage{graphicx}
\usepackage{epsfig}
\usepackage[centertags]{amsmath}
\usepackage{graphicx,indentfirst,amsmath,amsfonts,amssymb,amsthm,newlfont}
\usepackage{longtable}
\usepackage[usenames,dvipsnames]{xcolor}
\usepackage{url}
\usepackage{amsthm}
\usepackage{float} 

\newtheorem{defin}{Definição}

\begin{document}

\title{Uma avaliação rigorosa da intermitência no mapa logístico por meio do limite inferior do erro}

\author
    { \Large{Marcella Nathália Resende de Oliveira}\thanks{marcella.oliveeira@hotmail.com} \\
  \Large{Erivelton Geraldo Nepomuceno}\thanks{nepumuceno@ufsj.edu.br} \\
     {\small Departamento Engenharia Elétrica, UFSJ, São João del-Rei, MG,\\
     Grupo de Controle e Modelagem, GCOM}
   }

\criartitulo


\begin{abstract}
{\bf Resumo}. Este artigo investiga o tempo máximo de simulação em que o fenômeno da intermitência pode ser observado com confiança numérica em mapas discretos. Foram empregados conceitos de análise intervalar e o limite inferior do erro. Como resultado, foi observado que a confiabilidade da intermitência é dependente da condição inicial. Quatro exemplos numéricos mostram a eficiência da proposta.

\noindent
{\bf Palavras-chave}. Mapa logístico, Intermitência, Aritmética Intervalar, Limite inferior do erro.
\end{abstract}


\section{Introdução}

Sistemas dinâmicos são modelos matemáticos para muitos problemas na física, biologia, economia e engenharia. Dentre os inúmeros tipos de sistemas, os sistemas caóticos evoluem no domínio do tempo com um comportamento aperiódico, em que seu estado futuro é extremamente dependente do estado atual \cite{monteiro}. 

Uma parcela significativa da investigação sobre sistemas dinâmicos é realizada por meio de simulações computacionais. Hoje, existe a percepção de que se vive uma era cuja capacidade computacional é ilimitada, podendo atingir precisões arbitrárias. A realidade, porém, não é bem assim. A limitação de memória se torna um empecilho para o cálculo com precisão infinita, mesmo com o uso de si\-mu\-la\-ção simbólica. Conforme indicado em \cite{lozi}, existem muitos trabalhos publicados em que a confiabilidade dos resultados numéricos não é cuidadosamente verificada. Na investigação de alguns destes problemas, Lorenz \cite{lorenz} cunhou o termo ``caos computacional" enquanto estudava o comportamento caótico de equações diferenciais usadas para aproximar um sistema contínuo representado por um conjunto de equações di\-fe\-ren\-ci\-ais quando o tamanho do passo é aumentado. Nepomuceno \cite{convergence} apresentou um estudo em um dos softwares mais utilizados na área de pesquisa de Identificação de Sistemas, o Matlab, utilizando precisão dupla, no qual uma sequência simples de iterações convergiu para a resposta errada. Em continuidade aos trabalhos dessa linha, Rodrigues Júnior e Nepomuceno \cite{RN2015} utilizam análise intervalar para reproduzir os mesmos resultados obtidos em \cite{convergence}, mostrando mais uma vez que o cálculo dos pontos fixos pode exigir cuidadosa atenção no que se refere à computação numérica.

O propósito deste estudo é trabalhar com sistemas que apresentam comportamento intermitente - ora regular, ora caótico \cite{theory,pomeau1980intermittent}. Boa parte da pesquisa que envolve intermitência faz uso de si\-mu\-la\-ção computacional para reproduzir resultados numéricos relevantes. Entretanto, pouca atenção tem sido dada para o tempo máximo de simulação em que o fenômeno da intermitência pode ser observado. Neste trabalho, a busca pela indicação desse tempo será guiada pelo limite inferior do erro $  $\cite{Nepomuceno2016}. Também será feita uma análise da influência da condição inicial na observação deste fenômeno.

O artigo está organizado da seguinte forma. Na Seção 2, apresenta-se os conceitos preliminares a respeito de função recursiva e mapa logístico, o limite inferior do erro (lower bound error) e ponto fixo. Em seguida, na Seção 3, a metodologia é apresentada. Os resultados são apresentados na Seção 4, enquanto as considerações finais são indicadas na Seção 5.


\section{Conceitos Preliminares}
\subsection{Mapa Logístico}
Inicialmente, as funções recursivas podem ser definidas da seguinte forma, de acordo com \cite{convergence}: seja $\mathbf{I} \subseteq \mathbf{R}$ um espaço métrico com $f:\mathbf{I} \longrightarrow \mathbf{R}$ tem-se que:
\begin{equation}
x_n = f(x_{n-1}).
\end{equation}

O mapa logístico foi descrito pelo biólogo May \cite{May}. É uma equação que ao ter seus parâmetros variados, apresenta um comportamento diferenciado. O Mapa Logístico foi descrito como:
\begin{equation}
\label{eq:1a}
x_{n+1}=rx_n(1-x_n).
\end{equation}

A Equação \eqref{eq:1a} foi desenvolvida como um modelo populacional, com $x_n$ sendo
um número entre $0$ e $1$ que representa a razão entre a po\-pu\-la\-ção existente na n-ésima
geração e o maior número possível de indivíduos e $r$ como sendo uma taxa de crescimento
da população. Escolhendo um valor para o parâmetro $r$ e iterando recursivamente o mapa
a partir de uma condição inicial $x_0$, obtém-se uma série temporal da equação do mapa
logístico.

Trata-se, então, de um exemplo de função recursiva, capaz de reproduzir o comportamento de fenômenos
não-lineares. A Equação \eqref{eq:1a} chama atenção por ser um modelo bastante simples, mas capaz de se
comportar de maneira complexa. É extremamente sensível mesmo com pequenas variações das condições iniciais ou do número de iterações tomadas. Dessa forma, os erros são inerentes ao processo quando procura-se conhecer o comportamento para infinitas iterações.

\subsection{Ponto Fixo}

A partir da iteração subsequente de (1), é possível gerar séries de tempo discreto. A escolha de $f$ define o comportamento da série gerada: ponto fixo de período 1 ou maior ou comportamento caótico.
Segundo \cite{Fei}, se $f(x^*) = x^*$, então considera-se que $x^*$ é um ponto fixo de $f(x)$. O princípio de mapeamento da contração é um meio simples para encontrar o ponto fixo a partir de uma condição inicial com um número $x_0$ arbitrário e da definição da sequência ${x_n}$ por
$x_n = f(x_{n-1})$ \cite{rudin,ferrar}. Se essa sequência for convergente, então $x_n \rightarrow x^*$ à medida em que $n \rightarrow \infty$.

\subsection{Limite Inferior do Erro}

O limite inferior do erro (lower bound error) será utilizado para observação do erro inerente à simulação. Isso será um indicativo de que a simulação pode gerar intermitência de modo equivocado. Será indicado na metodologia que uma forma de ter certeza disso é quando utiliza-se $x_0 = 1/r$. O resultado final deveria ser o ponto fixo, mas por fim gera intermitência.

\begin{defin}
Sejam duas pseudo-órbitas ${\hat{x}_{a,n}}$ e ${\hat{x}_{b,n}}$ derivadas de duas extensões intervalares. 
\begin{equation}
    \delta_{\alpha,n}=\frac{|{\hat{x}_{a,n}}-{\hat{x}_{b,n}}|}{2}
\end{equation}
é o limite inferior do erro do mapa $f(x)$ quando $\delta_{a,n} \geq \delta_{\alpha,n}$ ou $\delta_{b,n} \geq \delta_{\alpha,n}$,
\end{defin}
em que $\delta_{a,n}$ e $\delta_{b,n}$ são os erros referentes a cada pseudo-órbita.

Considerando a equação do mapa logístico,
foram obtidas duas pseudo-órbitas derivadas da Equação \eqref{eq:1a} - apresentadas a seguir. Elas são equações matematicamente equivalentes, mas diferentes do ponto de vista da representação em ponto flutuante. Isso significa que duas sequências matematicamente equivalentes de operações aritméticas podem levar a dois resultados diferentes, devido às propriedades da aritmética real que não são totalmente válidas na aritmética de ponto flutuante \cite{Ove2001,IEEE2008}.
\begin{equation}
    x_a(k+1)=r \cdot x_a(k)-r \cdot (x_a(k))^2;
\end{equation}
\begin{equation}
    x_b(k+1)=r \cdot x_b(k) \cdot (1-x_b(k)).
\end{equation}





\section{Metodologia}

Dada a equação $x_{n+1}=rx_n(1-x_n)$, que descreve o mapa logístico, é possível observar o crescimento do erro na sua simulação e avaliar rigorosamente se é possível ou não garantir a existência de intermitência com a precisão computacional utilizada. Os passos para isso são:
\begin{enumerate}
    \item Determinação das extensões intervalares - Equações (4) e (5);
    \item Determinação do conjunto de parâmetros $r$ e condições iniciais a serem avaliados. Neste caso será utilizado o parâmetro $r=3,8283$ e quatro condições iniciais diferentes $x_0=0,3$ \cite{monteiro}, $x_0=1/r$, $x_0=300/341$ e $x_0=1904/6365$. Foi feita uma busca heurística por condições iniciais que pudessem comprovar a existência de regimes caóticos e regulares em uma mesma janela de tempo;
    \item Cálculo do limite inferior do erro:
    \begin{equation*}
        \delta_{\alpha,n}=\frac{|{\hat{x}_{a,n}}-{\hat{x}_{b,n}}|}{2};
    \end{equation*}
    \item Análise qualitativa das duas pseudo-órbitas;
    \item Cálculo do tempo máximo de simulação.
\end{enumerate}

A partir deste procedimento espera-se determinar até que ponto existe ou não intermitência e até quantas iterações isso ocorre.


\section{Resultados}
\begin{figure}[ht]
	\centering
	\begin{minipage}[b]{0.48\textwidth}
		\includegraphics[width=\textwidth]{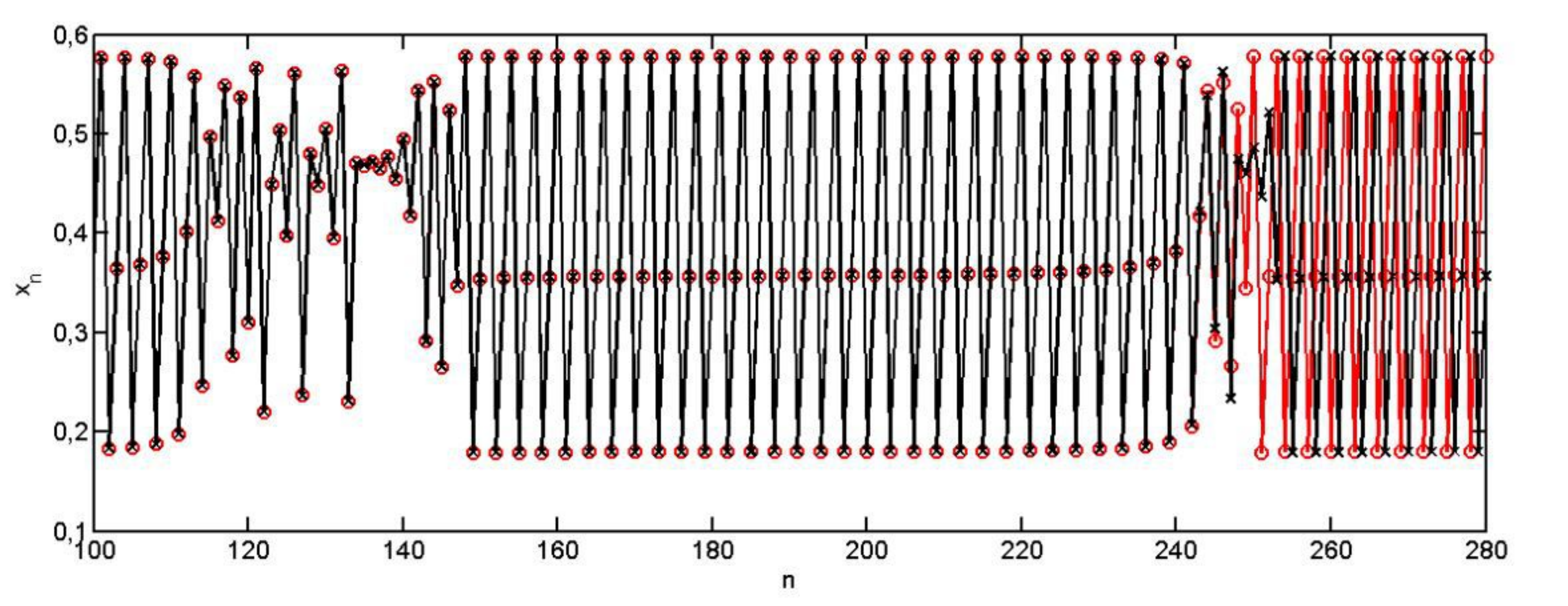}
	\caption{\label{fig:error1} Simulação de (2), com parâmetro $r = 3,8283$ e $x_0 = 0,3$.}	
	\end{minipage}
	\hfill
	\begin{minipage}[b]{0.48\textwidth}
		\includegraphics[width=\textwidth]{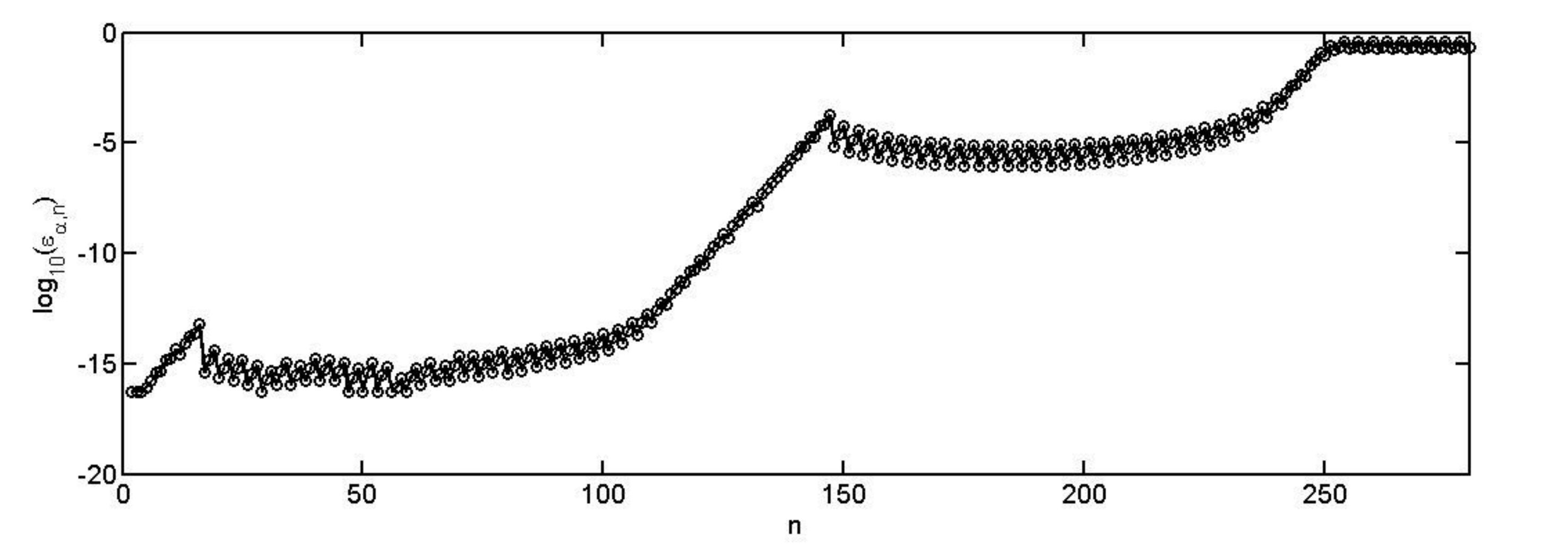}
\caption{\label{fig:error2} Evolução do erro da simulação de (2), com parâmetro $r = 3,8283$ e $x_0 = 0,3$.}
	\end{minipage}
\end{figure}

Os resultados obtidos com a simulação do mapa logístico para o parâmetro $r = 3,8283$ e a condição inicial $x_0 = 0,3$ são apresentados nas Figuras \ref{fig:error1} e \ref{fig:error2}. 

\begin{figure}[ht]
	\centering
	\begin{minipage}[b]{0.48\textwidth}
		\includegraphics[width=\textwidth]{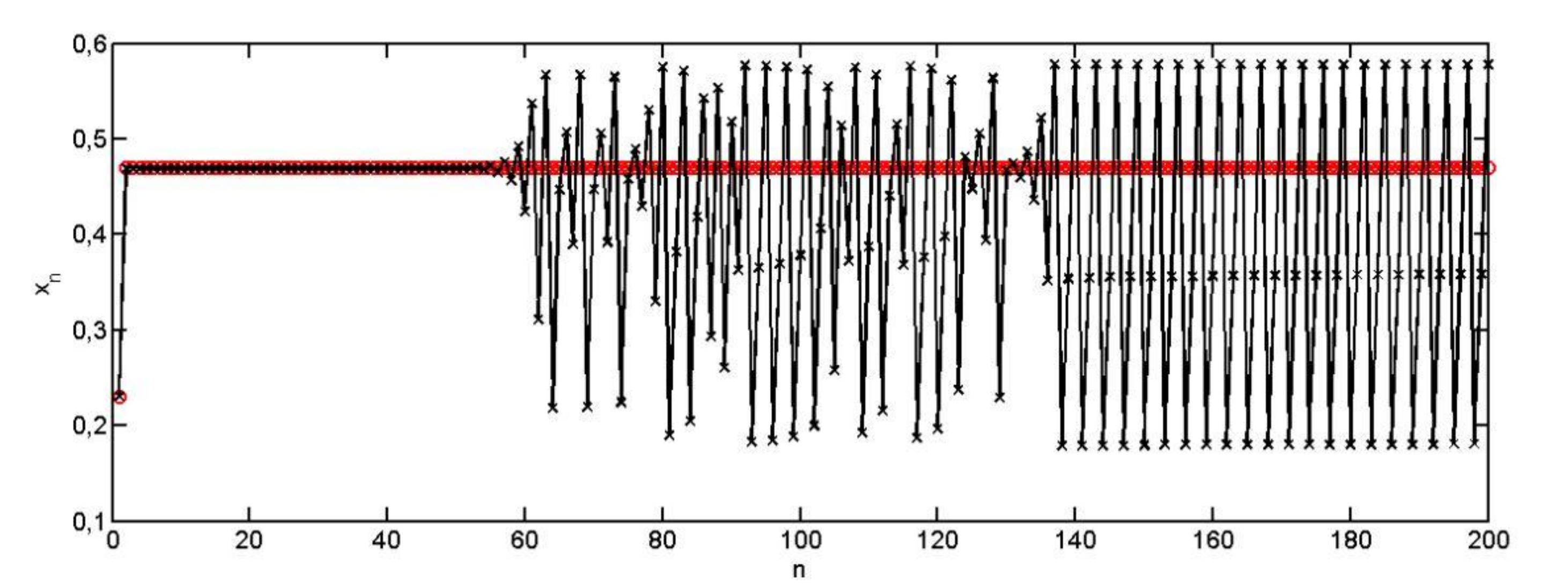}
\caption{\label{fig:error1a} Simulação de (2), com parâmetro $r = 3,8283$ e $x_0 = 1/r$.}
	\end{minipage}
	\hfill
	\begin{minipage}[b]{0.48\textwidth}
	\includegraphics[width=\textwidth]{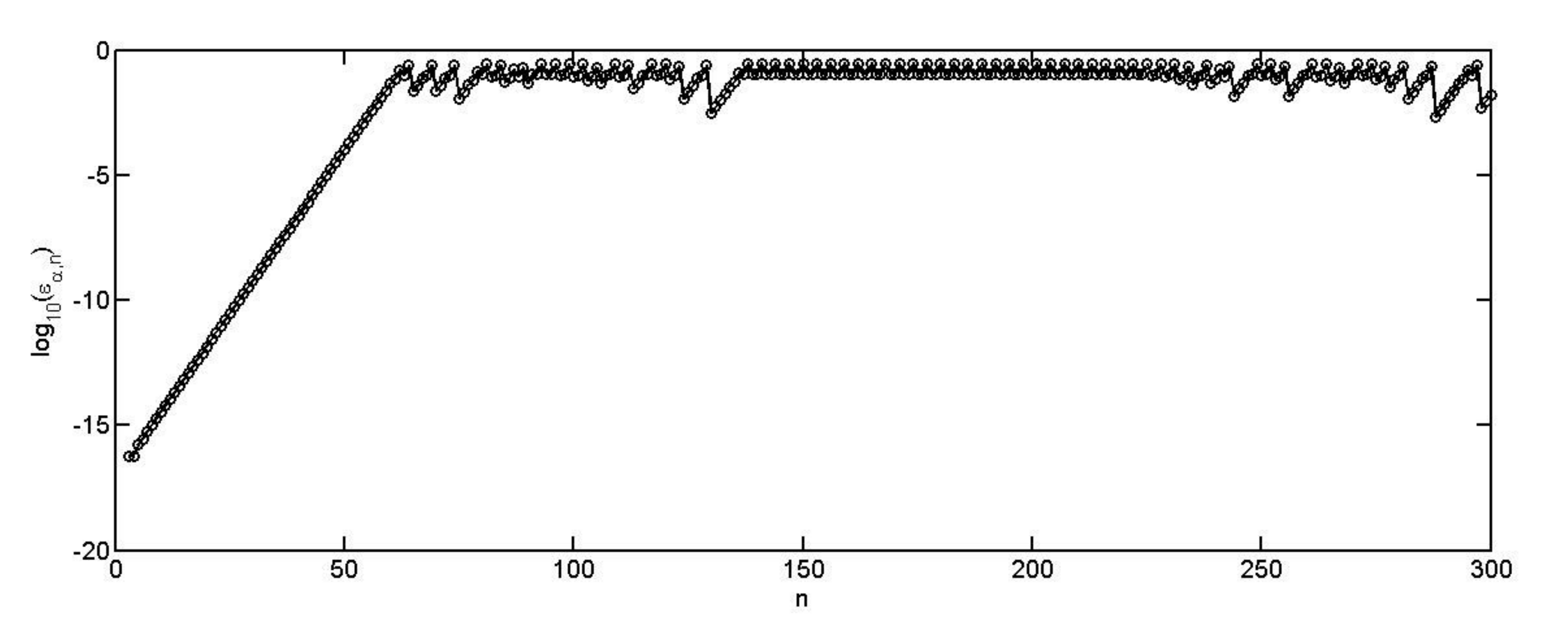}
\caption{\label{fig:error2a} Evolução do erro da simulação de (2), com parâmetro $r = 3,8283$ e $x_0 = 1/r$.}	
	\end{minipage}
\end{figure}

Os resultados obtidos com a simulação do mapa logístico para o parâmetro $r = 3,8283$ e a condição inicial $x_0 = 1/r$ são apresentados nas Figuras \ref{fig:error1a} e \ref{fig:error2a}. 

\begin{figure}[ht]
	\centering
	\begin{minipage}[b]{0.48\textwidth}
		\includegraphics[width=\textwidth]{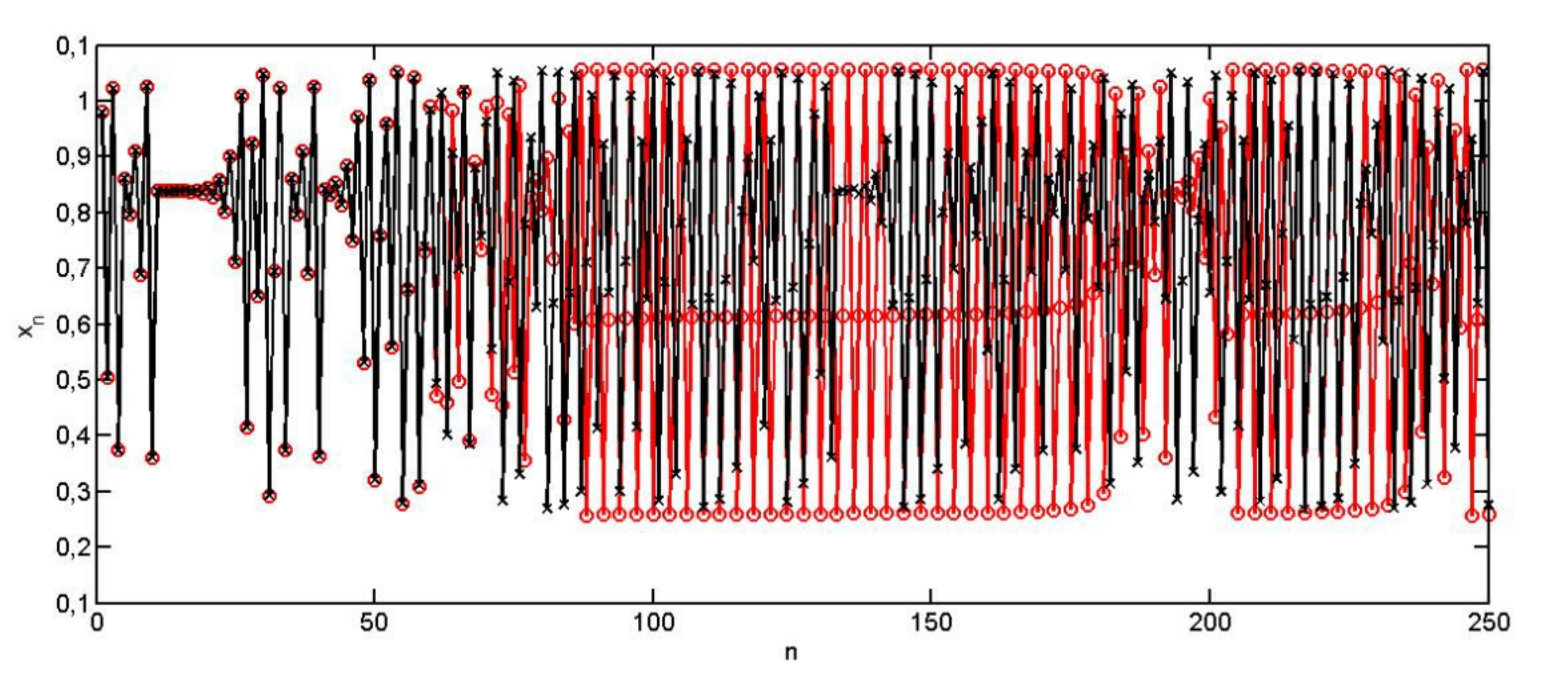}
\caption{\label{fig:error1b} Simulação de (2), com parâmetro $r = 3,8283$ e $x_0 = 300/341$.}
	\end{minipage}
	\hfill
	\begin{minipage}[b]{0.48\textwidth}
		\includegraphics[width=\textwidth]{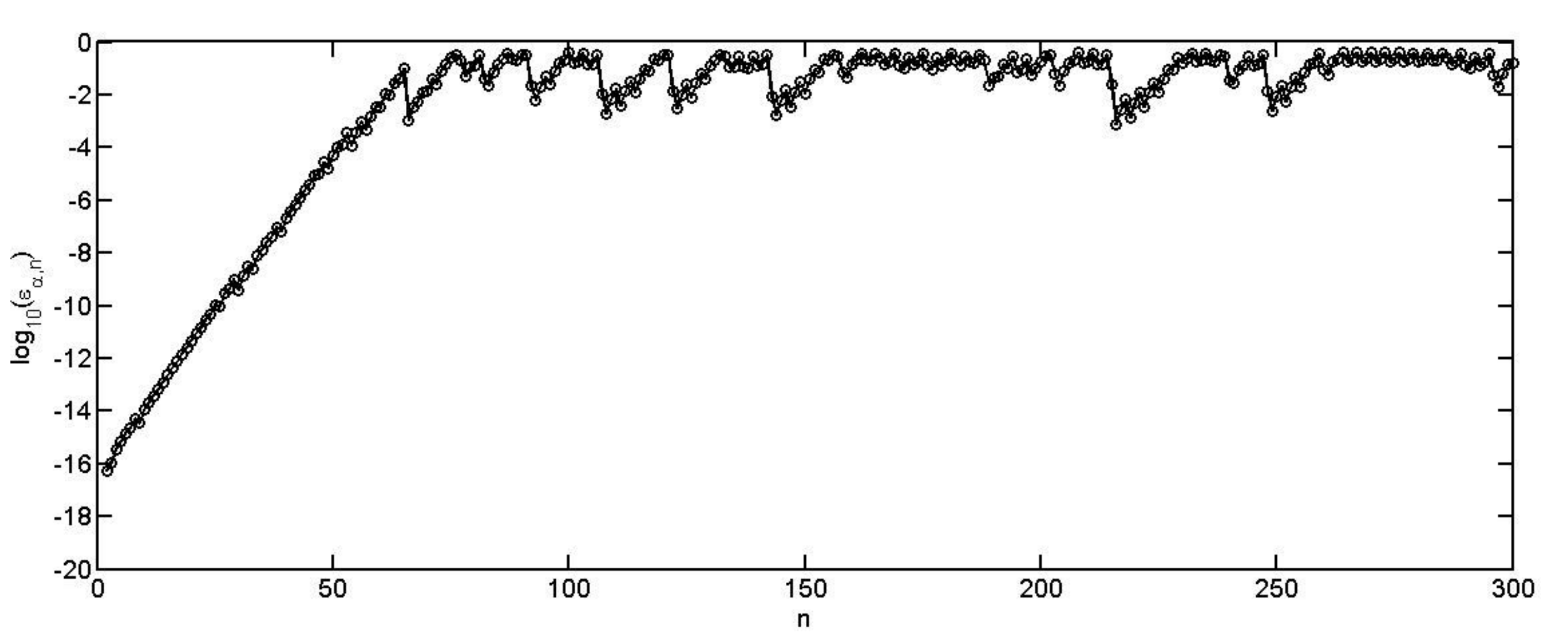}
\caption{\label{fig:error2b} Evolução do erro da simulação de (2), com parâmetro $r = 3,8283$ e $x_0 = 300/341$.}	
	\end{minipage}
\end{figure}

Os resultados obtidos com a simulação do mapa logístico para o parâmetro $r = 3,8283$ e a condição inicial $x_0 = 300/341$ são apresentados nas Figuras \ref{fig:error1b} e \ref{fig:error2b}. É observado que logo no primeiro regime laminar (regular) para uma das extensões intervalares, a outra ainda continua com comportamento caótico.

\begin{figure}[ht]
	\centering
	\begin{minipage}[b]{0.48\textwidth}
		\includegraphics[width=\textwidth]{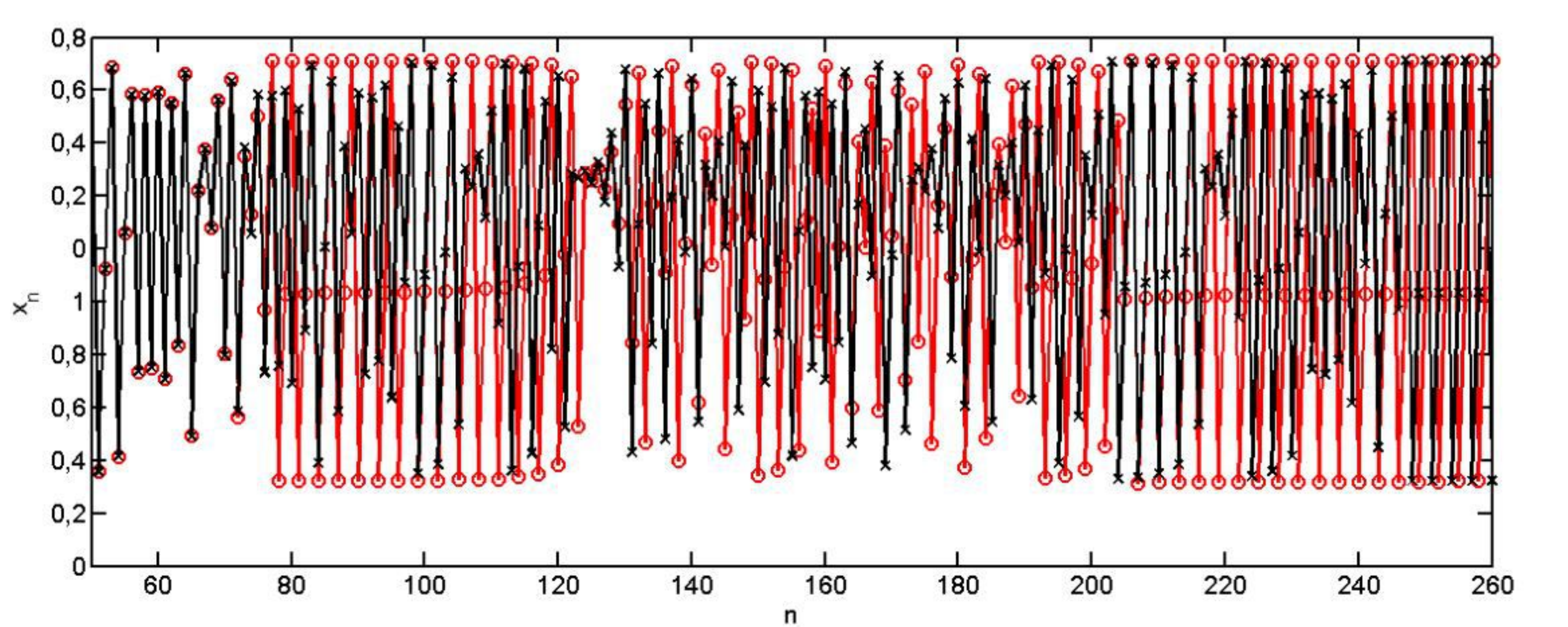}
\caption{\label{fig:error1c} Simulação de (2), com parâmetro $r = 3,8283$ e $x_0 = 1904/6365$.}
	\end{minipage}
	\hfill
	\begin{minipage}[b]{0.48\textwidth}
		\includegraphics[width=\textwidth]{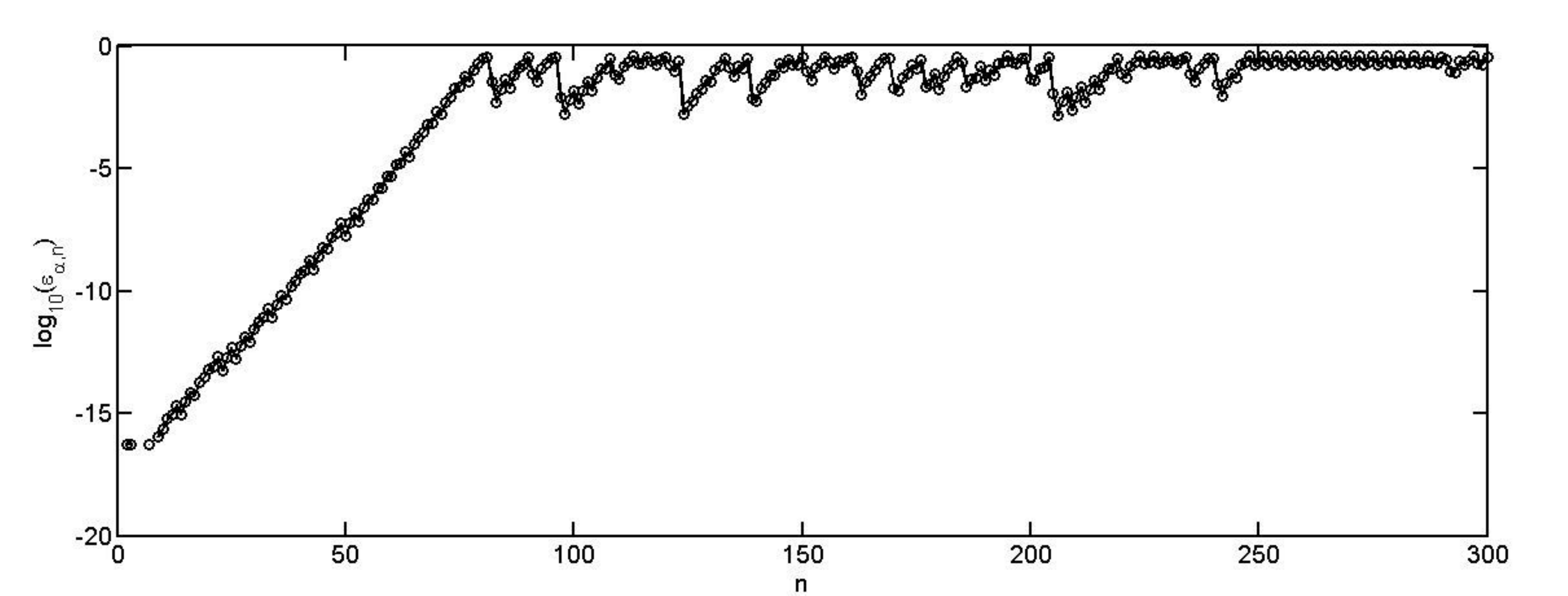}
\caption{\label{fig:error2c} Evolução do erro da simulação de (2), com parâmetro $r = 3,8283$ e $x_0 = 1904/6365$.}	
	\end{minipage}
\end{figure}

Os resultados obtidos com a simulação do mapa logístico para o parâmetro $r = 3,8283$ e a condição inicial $x_0 = 1904/6365$ são apresentados nas Figuras \ref{fig:error1c} e \ref{fig:error2c}. É observado que logo no primeiro regime laminar (regular) para uma das extensões intervalares, a outra ainda continua com comportamento caótico. O mesmo fato foi observado na Figura \ref{fig:error1b}. 

	Pela análise das simulações, é possível observar que o comportamento intermitente é dependente de $x_0$, ou seja, ele se apresenta de di\-fe\-ren\-tes formas quando a condição inicial é mo\-di\-fi\-ca\-da. Além disso, percebe-se que ele se apresenta na forma de erro numérico ou inconsistência matemática quando a interseção entre os intervalos de iterações consecutivas é diferente de conjunto vazio e, dessa forma, não se pode afirmar que são resultados diferentes. O tempo máximo de simulação em que o fenômeno da intermitência pode ser observado está relacionado à ocorrência de perda de dígitos significativos ao longo das iterações e geração de resultados providos de erro \cite{Gol1991,Ove2001}. Assim, uma vez que a simulação de equações matemáticas equivalentes gera resultados alternadamente laminares e caóticos, a confiabilidade da simulação é afetada.


\section{Conclusão}
De fato, como foi visto, a condição inicial $x_0$ é um elemento que influencia consideravelmente a presença de intermitência no sistema. Além disso, o tempo máximo de simulação em que o fenômeno da intermitência pode ser observado está relacionado à ocorrência de perda de dígitos significativos ao longo das iterações e geração de resultados providos de erro. Assim, uma vez que a simulação de equações matemáticas equivalentes gera resultados alternadamente laminares e caóticos, não se pode afirmar que são resultados diferentes e a confiabilidade da simulação é afetada.

Ao longo dos processos históricos da ciência e da engenharia, a computação numérica recebe grande atenção por ter se tornado uma poderosa ferramenta para resolução numérica de problemas matemáticos. Por outro lado, apesar de sua importância para a infraestrutura científica moderna e de apresentar resultados satisfatórios e muito próximos dos esperados, a computação numérica ainda está longe de ser uma ferramenta que disponibiliza resultados totalmente de acordo com a realidade. Isso acontece devido à limitação de memória da máquina, que se torna um empecilho para o cálculo com precisão infinita e, consequentemente, interfere na resposta do sistema. Dessa forma, a comprovação de resultados obtidos por meio da aritmética computacional, perpassa, antes de tudo, o entendimento sobre como o computador lida com as operações matemáticas e arredondamentos.


\section{Agradecimentos}

Agradecemos ao CNPq/INERGE, à Universidade Federal de São
João del-Rei e aos membros do grupo de controle e modelagem (GCOM) pelo apoio.
\\
\bibliographystyle{plain}
\bibliography{ref}
\end{document}